# On Electrical Equivalence of Aperture-Body and Transmission-Cavity Resonance Phenomena in Subwavelength Conducting Aperture Systems from an Equivalent Circuit Point of View

Young-Ki Cho[1], Jong-Ig Lee[2], and Ki Young Kim[3,*]

[1]*School of Electrical Engineering and Computer Science, Kyungpook National University, Daegu 702-701, Korea*

[2]*Department of Electronics Engineering, Dongseo University, Pusan 617-716, Korea*

[3]*Research and Development Division for Hyundai Motor Company and Kia Motors Corporations, Hwaseong 445-706, Korea*

For a narrow slit structure backed by a conducting strip which is taken as a representative example of an aperture-body resonance (ABR) problem, the transmission resonance condition (i.e., condition for maximum power transmission) and the transmission width (i.e., normalized maximum transmitted power through the slit) are found to be the same as those for narrow slit coupling problem in a thick conducting screen, which is designated as a transmission-cavity resonance (TCR) problem. From a viewpoint of equivalent circuit representation for the transmission resonance condition and the funneling mechanism, the ABR and the TCR problems are thought to be essentially of the same nature.

---

[*] Author to whom correspondence should be addressed.

## 1. INTRODUCTION

The electromagnetic (EM) coupling problem through apertures – generally referred to as the aperture coupling problem – has a relatively long history in the area of EM-related research, as described in Ref. 1. With progress having been made in this area, the main focus of concern has shifted to problems of greater complexity, e.g., more complex structures such as transmission resonance problems through narrow slits or small apertures. In these problems, the transmission resonance condition under which the maximum power transmission through the slit or aperture occurs is mainly dealt with. The aperture-body resonance (ABR) problem[2] is representative of such transmission resonance problems. This problem is composed of a small aperture backed by a conducting body with the main interest focused on investigating the condition under which the transmitted power through the small aperture is increased significantly over that of the case in which the scatterer near to the small aperture is not present.

There is another aperture coupling problem which draws concern from the viewpoint of transmission resonance phenomena. This coupling problem presupposes a resonant transmission cavity instead of the nearby scatterer in the ABR problem. In this sense, such a coupling problem will be designated as the transmission-cavity resonance (TCR) problem in order to differentiate it from the ABR problem. A typical example of this type is the transmission resonance problem[3] through a narrow slit in a thick conducting screen.

Although these two problems, ABR and TCR, are seemingly different from each other, they have been observed as sharing common features in the light of the transmission resonance through the slit. We will therefore discuss such features between two problems from the generalized viewpoint of the transmission resonance associated with aperture coupling problems from an equivalent circuit point view. These two resonance problems, which have been extensively studied as microwave problems, could also provide basic underlying physics of nanoscale science and engineering associated with subwavelength apertures such as extraordinary optical transmission (EOT) through subwavelength aperture[4], near-field scanning optical microscope (NSOM) for nanoscale imaging applications[5], very small aperture laser (VSAL) system[6], optical antennas[7], and many more.

## 2. APERTURE-BODY RESONANCE (ABR) PROBLEM

As a simplified model of the ABR problem, the two-dimensional problem composed of a narrowly slitted conducting screen and a nearby conducting strip, as shown in Figure 1(a), is considered. In Figure 1(a), $\mu_0$ and $\varepsilon_0$ are the permeability and permittivity of the free space, respectively. $P_{inc}$ and $P_t$ are the incident power density and transmitted power, respectively. $E_{inc}$ ($H_{inc}$) is the electric (magnetic) field incident on the slit with incidence angle $\theta_0$ with respect to the z-axis, $H_0$ is the amplitude of the incident magnetic field, and $k_0 (= 2\pi/\lambda_0, \lambda_0$ ; free space wavelength) is the propagation constant in the free space. $a$ is the width of the slit and $d$ is the distance between slitted conducting screen and the nearby conducting strip. The equivalent circuit representation for this structure with narrow slit can be given in Figure 1(b). $I_S$ is the equivalent source current corresponding to sum of the incident and reflected fields in Figure 1(a).

For this analysis, the equivalence principle is employed in order to divide the original problem into two equivalent situations by placing appropriate equivalent magnetic currents over both sides of the shorted slit. Pair of coupled integral equations, whose unknowns are the magnetic current density over the shorted slit and the induced electric current density on the conducting strip, are set up. Next, the pulse basis and Galerkin scheme for the unknown magnetic current densities over the slit, together with the piecewise sinusoidal Galerkin scheme for the unknown induced current densities on the strip, are employed in order to reduce the aforementioned coupled integral equations to a linear equation system. Accordingly, based on knowledge of the unknown current densities, the quantities of interest are obtained, including the coupled (transmitted) power $P_t$ through the slit into the region 2, transmission width or apparent width $a_{tw}$ of the slit, and the equivalent circuit parameters of Figure 1(b) for the input admittances, $Y_1$ and $Y_2$, as seen by the slit looking into the regions 1 and 2, respectively, at the slit. For further details, refer to Ref. 8 to 11.

When the problem is specialized to the case of a narrow slit, i.e., $k_0 a \ll 1$, with no nearby conducting strip present, the admittances of Figure 1(b), $Y_1$ and $Y_2$, looking into the left and right half spaces, respectively, are the same as

$$Y_1 = Y_2 = G + jB = \frac{k_0}{2\eta_0}\left[1 - j\frac{2}{\pi}\log(k_0 aC)\right] \tag{1}$$

in which $C = 0.1987$.[3] Note that the conductance $G\,(= k_0/2\eta_0 = 1/120\lambda_0\,[\text{S/m}])$ is the typical value of conductance appearing in the various types of narrow slit problems, such as narrow and wide transverse slits [12, 13] in a parallel-plate waveguide (PPW) of small guide height, and a narrow slit in a flanged PPW,[14] where $\eta_0$ is the impedance of the free space and is given by $120\pi$. It is also worth mentioning that the susceptance $B$ becomes approximately equal to that of a narrow slit in a flanged PPW of small guide height. This is because, in the latter case, most of the reactive power near the slit is confined to the exterior region, and so the reactive power is very small in the interior region near the flanged slit.[14] In the case of a narrow slit with no nearby scatterer present, the conductance $G$ of the slit is much smaller than the susceptance, i.e., $G \ll B$. From the equivalent circuit in Figure 1(b), the transferred power $P_t$ to the load $Y_2$, corresponding to the transmitted power through the slit into the region 2, is given as

$$P_t = \frac{1}{2}\text{Re}\{V_s I_2^*\} = \frac{1}{2}\frac{H_0^2 G}{G^2 + B^2} \quad [\text{W/m}] \tag{2}$$

which is seen as being very small in the case of a narrow slit.

The power transmitted by an electrically narrow slit backed by a nearby conducting strip, however, can be much larger than that when the strip is absent. In the equivalent circuit for a narrow slit, when the strip is present in region 2, the admittance $Y_2$ would be changed, whereas the admittance $Y_1$ remains same, irrespective of the presence of the strip in region 2. Maximum power transmission and its condition have been searched by inspecting the transmitted power $P_t$ for various combinations of the strip length $L$ and the location ($X_0, d$) of the nearby strip center with a narrow slit width $a$ kept unchanged. When the maximum power transmission occurs, the input admittance $Y_2$ looking into the region 2 is found to be complex conjugate of $Y_1$, i.e., $Y_2 = Y_1^* = G - jB_1$. As a result, the total admittance $Y_S (= Y_1 + Y_2 = 2G = 1/60\lambda_0$ [S/m]) becomes real value. This corresponds to the ABR condition, under which the maximum transmission power $P_{tm}$ through the slit is obtained as $H_0^2 / 2G = 60\lambda_0 H_0^2$ [W/m] by use of the simple circuit theory in Figure 1(b). Note that $P_{tm}$ has the physical dimension of [W/m], i.e., transmitted power per unit length along the $y$-axis, whereas the projected component of the incident power density $P_{inc}$ on the screen for an arbitrary incidence angle $\theta_0$ is given by

$$P_{inc} = \frac{1}{2}\eta_0 H_0^2 \cos\theta_0 \ [W/m^2]. \tag{3}$$

Therefore, if we define the ratio of $P_t / P_{inc}$ so that the ratio has a dimension of length, this ratio $a_{tw} (= P_t / P_{inc})$ [m] can have a physical meaning of transmission width or apparent width of the slit along the $x$-axis transverse to the slit axis. In particular, for normal incidence (i.e., $\theta_0 = 0$), this ratio becomes a maximum of $\lambda_0 / \pi$ [m], which, interestingly enough, corresponds to the effective height of a half wavelength dipole antenna.[15]

This can be physically interpreted as that incident wave power upon the apparent width $\lambda_0 / \pi$ [m] of the slit much larger than the actual slit width $a$ is funneled[16] into the slit and radiated into the right half space under the ABR condition. As a result of this, the power transmission through the slit is remarkably enhanced, regardless of the actual slit width.

We have investigated the variation of the transmission width $a_{tw}$ of the slit against the lateral strip offset $X_0$ for sets of $\{L, X_0, d\}$ found for the maximum transmission under the ABR condition. Figure 2 shows two contrastive types of the ABR phenomena. The curve of solid line shows two sharp peaks of the power transmission at the two offset positions, $X_0/\lambda_0 \cong \pm 0.24$. This means that the variation of $a_{tw}$ is very sensitive to the offset $X_0$. On the other hand, the curve of the dashed line reaches its maximum when the offset is zero. In addition, this curve at maximum is insensitive to the offset near the zero offset position $X_0 = 0$. Note that the maximum values of $a_{tw} (= \lambda_0/\pi)$ for both the former and the latter cases are the same. For the former case, the cavity mode field is strongly excited underneath the strip as in a normal rectangular microstrip antenna structure. This type of maximum power transmission only occurs when the separation $d$ between the strip and the conducting screen is so small that an electrical cavity is formed underneath the strip. Accordingly, under the condition of maximum power transmission, the former type can be regarded as an extreme case in which the separation $d$ is so small that a strong resonant mode field is set up underneath the strip, as mentioned above. Conversely, the latter type of maximum power transmission corresponds to the other extreme case and only occurs when the values of $d$ are considerably larger than those in the former case. Although the two cases of maximum transmission are different from each other, the role of the strip is the same in that its presence provides the required inductive susceptance for transmission resonance; in other words, $Y_2$ is made to be $Y_1^*$, i.e., $Y_2 = Y_1^*$. These two kinds of coupling are similar to those in the prior work on the slit-to-strip coupling in the PPW structure.[9]

## 3. TRANSMISSION-CAVITY RESONANCE (TCR) PROBLEM

The narrow slit coupling problem in a thick conducting screen, as shown in Figure 3, exemplifies the TCR problem, which presupposes a transmission cavity. In case of the structure in Figure 3(a), the gap region ($|x| \leq a/2, 0 \leq z \leq d$) constitutes a transmission resonant cavity, as seen from the equivalent circuit represented in Figure 3(b). There, the inside of the gap region is represented by the transmission line of PPW, and the flanged aperture at both ends of the gap region is represented by the radiation admittance $Y_L (= G + jB_L, G = 1/120\lambda_0$ [S/m]). For this problem, the slit width $a$ is assumed to be much smaller than the free space wavelength $\lambda_0$ so that only the transverse electromagnetic (TEM) mode is guidable.[17] Therefore, the characteristic admittance $Y_C$ is expressed by $Y_C = 1/\eta_0 a$. Note that $Y_C >> |Y_L|$, $B_L >> G$, and $G = 1/120\lambda_0$ [S/m] for $a << \lambda_0$, as discussed in Ref. 12 and 13. In this case, the reflection coefficient at the aperture plane is roughly close to unity. This physical situation corresponds to the lossy magnetic wall. This is essentially the same as the magnetic wall concept in the cavity model of the microstrip antenna theory.[18] Therefore, the present structure in itself constitutes the lossy resonant cavity. It should, however, be remembered that the structure for the previous ABR presupposes the parasitic scatterer near the slit, whereas the present structure presupposes the lossy resonant cavity.

In addition, for the TCR problem, the transmitted power $P_t$ through the slit into the region 2 reaches its maximum when the total admittance $Y_S$ becomes a real value, that is, when the admittance $Y_L'$ (obtained after transforming $Y_L$ at $z = d$ along the transmission line of the PPW region to the left input port at $z = 0$) becomes the complex conjugate of $Y_L$, i.e., $Y_L' = Y_L^* = G - jB_L$, and so $Y_L$ at $z = 0$ and $Y_L'$ are summed to be purely real, i.e., $Y_S = 2\text{Re}\{Y_L\} = 2G$. Under this condition, the transmission width $a_{tw}$ of the slit becomes $\lambda_0/\pi$ [m] for normal incidence cases ($\theta_0 = 0$) regardless of the actual slit width $a$. Therefore, the transmitted power through the slit is considerably increased by way of the funneling mechanism, which is the same as that for the ABR problem.

## 4. DISCUSSION AND CONCLUSION

We have discussed common features between two aperture coupling problems: ABR and TCR problems.

Based on the foregoing discussion, the common features between them are summarized as follows:

First, the maximum power transmissions occur when the total admittance at the input port of the incident side becomes real, i.e., $Y_S = 2G$ for both problems. It should be noted that $G(=1/120\lambda_0\,[\text{S/m}])$ refers to the radiation conductance seen by the slit into the single half space region.

Second, based upon the above observation, it can be said that the nearby scatterer in ABR and the resonant cavity in TCR play the same role in that they provide the required inductive susceptance for resonances, i.e., cancellation of the reactive component. In this sense, the two problems, though seemingly different from each other, can be taken as a general transmission resonance problem associated with aperture coupling problems as a unified viewpoint.

Third, under the transmission resonance condition, the transmission width of the slit $a_{tw} = \lambda_0/\pi\,[\text{m}]$ is identical for normal incidence cases, which, interestingly enough, correspond to the effective height of the half wavelength dipole antenna. The observation that the transmission width $a_{tw}$ of the slit can be much larger than the actual slit width leads us to the concept of "funneling mechanism", via which the power transmission through even the very narrow slit can be considerably enhanced.

It is also worth noting that the transmission resonance condition $Y_S = 2G$ for both ABR and TCR problems is the same as the maximum radiation condition $Y_S = 2G$ for the input admittance at the radiating edge in the transmission line model of the rectangular microstrip patch antenna.[19] Here $G$ means the conductance of the radiating edge of the rectangular microstrip patch.

The result that, at transmission resonance, the transmission width $a_{tw}(=\lambda_0/\pi\,[\text{m}])$ of the actual narrow slit is identical for both ABR and TCR problems regardless of the actual slit width is not unexpected, since similar results have been obtained in other transmission problems. For example, the transmission area of a small resonated aperture was obtained to be $3\lambda_0^2/4\pi\,[m^2]$ independent of the size or shape of the small aperture for the case of the small aperture in a conducting screen loaded by a lumped capacitor across its midpoint.[2] This can be taken as the three-dimensional structure of the aperture body resonance in which the nearby scatterer is replaced by a lumped element. In this structure, the lumped element plays the same role as the nearby scatterer in that the presence of both the two provides the required susceptance for resonance (i.e., cancellation of reactive component which the small aperture alone has). Moreover, the transmission width of the aperture-cavity-aperture system[20] was also found to be the same as that for the above small resonated aperture. This transmission problem of the aperture-cavity-aperture system can be taken as a three-dimensional TCR problem. So these two previous results suggest that the transmission widths are the same also for the two-dimensional ABR and TCR problems. It deserves mentioning that the transmission cavity resonance is observed to occur also for the narrow slot structure in thick conducting screens,[21] if the guide height of the rectangular guide structure inside the thick screen is much smaller than the wavelength. In this case too, the transmission cavity is composed of the finite length of transmission line terminated by very small radiation admittance at both left and right aperture planes[21] just like the above two-dimensional TCR problem.[3]

Over the last decade, interest has grown in enhanced transmission through periodic metallic samples, such as hole arrays[22] and deep metallic gratings.[23, 24] While the original work by Ebbesen et al.[22] was strictly limited to holes, some model calculations have been undertaken for deep metallic gratings.[23, 24] In the case of the original work, most experts agree that the enhanced transmission is related to surface plasmon excitation, whereas discussion still continues about the physical reason for such transmission. On the other hand, for the transmission grating, the enhanced transmission is attributed to the two different mechanisms associated with the cavity modes and the excitation of surface plasmon. It is worth pointing out that the enhanced transmission associated with the cavity mode is essentially the same as the transmission resonance for the above TCR problem. However it remains to be investigated further whether the enhanced transmission associated with the surface plasmon can be explained in the same framework of the equivalent circuit representation as that in the present TCR problem. This study may provide an alternative view of several nanoscale light propagation applications adopting subwavelength metallic apertures as mentioned previously.

**ACKNOWLEDGMENT**

This work was supported in part by BK21.

**Figure captions**

**Figure 1**   (a) The ABR structure under consideration and (b) its equivalent circuit representation.

**Figure 2**   Normalized transmission width $a_{tw}/\lambda_0$ of the slit versus the normalized strip offset $(X_0/\lambda_0)$.

**Figure 3**   (a) The structure of the TCR problem and (b) its equivalent circuit representation.

**Figure 1**

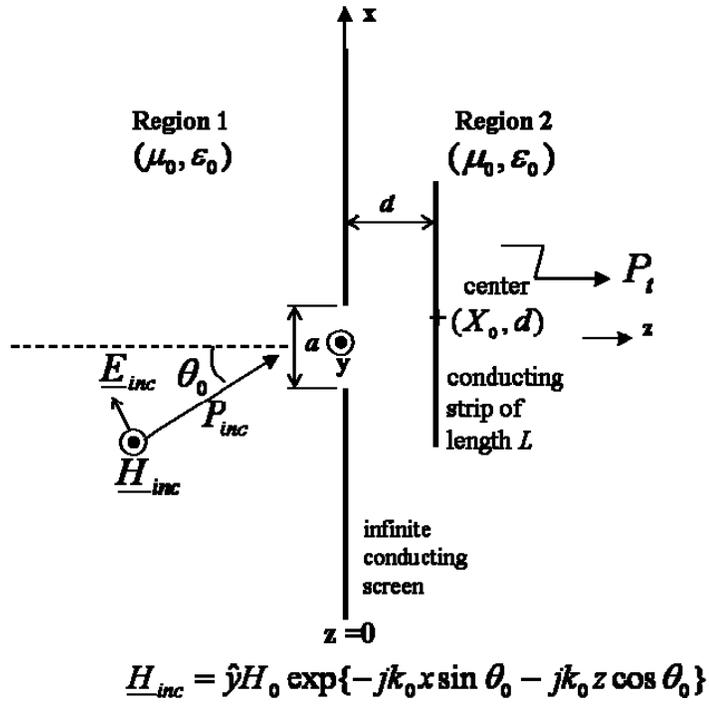

(a)

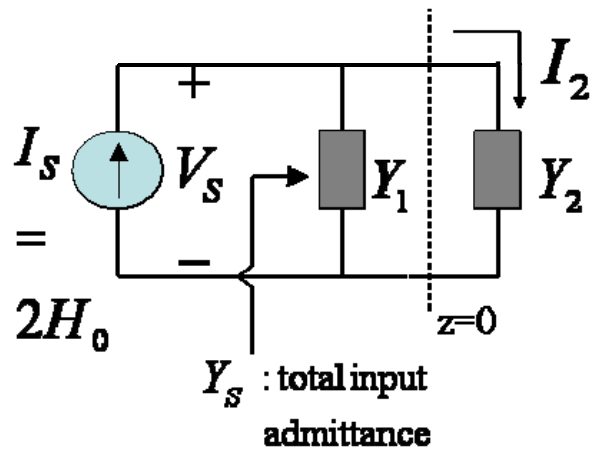

(b)

**Figure 2**

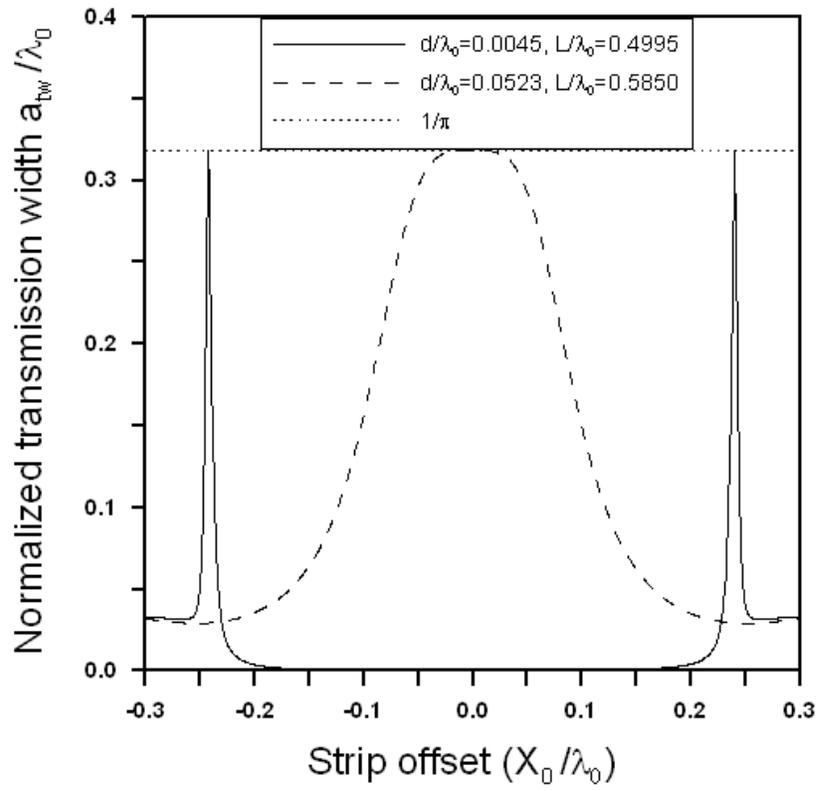

**Figure 3**

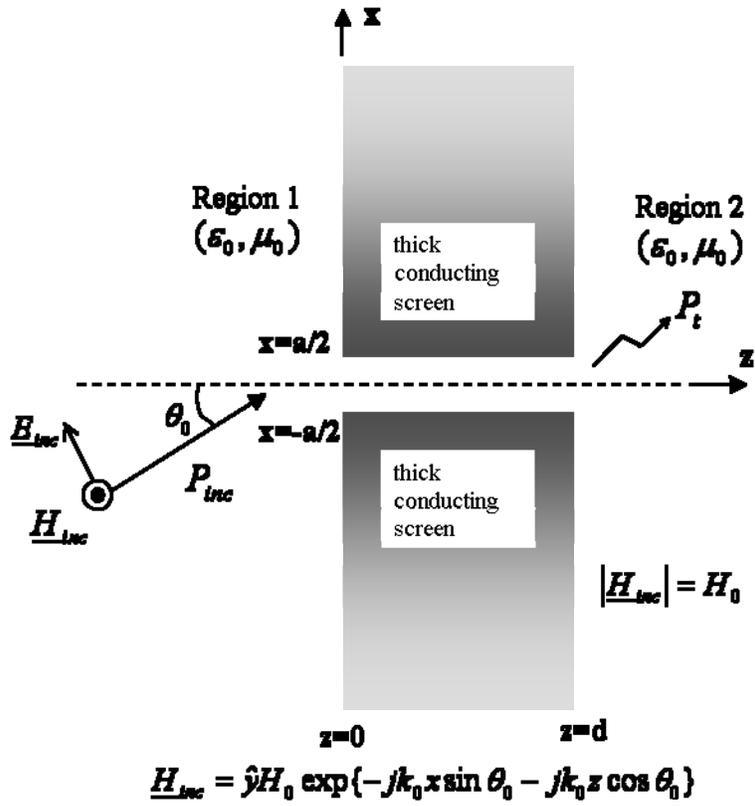

(a)

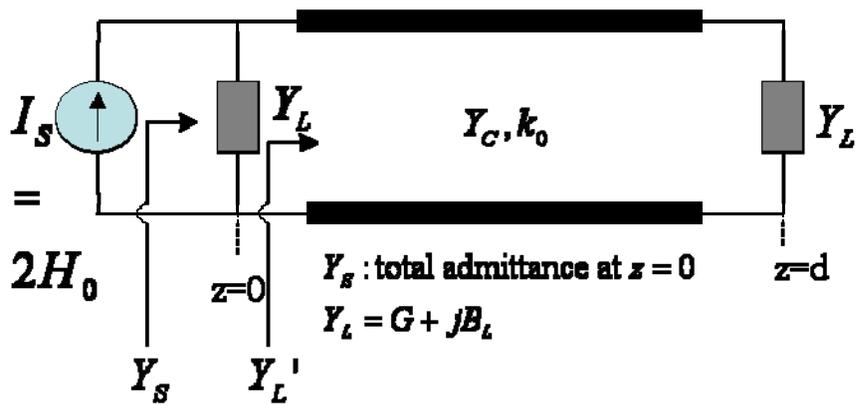

(b)